\documentclass[printer]{aa}

\bibpunct{(}{)}{;}{a}{}{,}

\usepackage{lscape}
\usepackage{graphicx}
\usepackage[varg]{txfonts}
\def\d{$^\circ$}
\def\m{$^\prime$}

\def\cm3{cm$^{-3}$}

\def\2{$^{12}$CO}
\def\3{$^{13}$CO}
\def\8{C$^{18}$O}

\def\cm2{cm$^{-2}$}

\begin{document}

\title{Radio and infrared study of the supernova remnant candidate HESS J1912$+$101}
\author { L. Duvidovich \inst{1,2}
\and A. Petriella \inst{1,3}
}

\institute{CONICET-Universidad de Buenos Aires, Instituto de Astronom\'ia y F\'isica del Espacio (IAFE), Buenos Aires, Argentina,
        \email{duvidovich@iafe.uba.ar}
\and Universidad de Buenos Aires, Facultad de Ciencias Exactas y Naturales, Buenos Aires, Argentina
\and Universidad de Buenos Aires, Ciclo B\'asico Com\'un, Buenos Aires, Argentina
  }

\offprints{L. Duvidovich}

   \date{Received <date>; Accepted <date>}

\abstract{}{We provide new insights into the $\gamma$-ray emission from HESS J1912$+$101, a TeV supernova remnant candidate probably
associated with the radio pulsar PSR J1913$+$1011.}
{We obtained new observations at $1.5\,\mathrm{GHz}$ using the VLA in the D configuration, with the 
purpose of detecting the radio shell of the putative remnant.  
In addition, we observed a single pointing at $6.0\,\mathrm{GHz}$ toward PSR J1913$+$1011 to look for a radio pulsar wind nebula. 
We also studied the properties of the surrounding interstellar medium using data of the $^{13}$CO, HI, and infrared emissions, obtained
from public surveys.}
{We did not find evidence of a radio shell down to the sensitivity of the new image at $1.5\,\mathrm{GHz}$.
We detect faint diffuse emission around PSR J1913$+$1011 at $6.0\,\mathrm{GHz}$, which could represent a radio pulsar wind nebula
powered by the pulsar. We find dense ambient gas at $\sim 60\,\mathrm{km\,s^{-1}}$, which shows a good spatial correspondence with 
the TeV emission only in the western and eastern directions. There is also dense gas near the center of HESS J1912$+$101, where
the TeV emission is weak. Using infrared data, we identify an active star-forming region in the western part of the shell.}
{Based on the poor spatial match between the ambient gas and the TeV emission (which shows a good correlation in the western and eastern directions and 
an anticorrelation in the other directions), we conclude that the hadronic mechanism alone does not give a satisfactory
explanation of the $\gamma$ rays from HESS J1912$+$101. Additional contributions may come from leptonic processes in the shell of the supernova remnant, together with contributions from PSR J1913$+$1011 and its pulsar wind nebula and/or from the star-forming region. 
A confident determination of the distance to the putative remnant is necessary to determine whether these sources are 
associated or just appear superimposed in the line of sight.  
}
{}

\titlerunning{Radio and infrared study of HESS J1912$+$101}
\authorrunning{L. Duvidovich et al.}


\maketitle

\section{Introduction}
\label{introd}

The population of very high-energy sources with a shell appearance has greatly increased in recent years. 
Based on the morphological correspondence with nonthermal emission in the radio and/or X-ray bands,  
many of them have been confirmed as supernova remnants (SNRs). Examples of this include SN 1006, Vela Jr, G347.3$-$0.5, RCW 86, and G353.6$-$0.7. 
Despite the fact that in these sources a SNR origin for the TeV emission is uncontroversial, in some cases the 
emission mechanism that produces very high-energy photons has not been firmly established. 
Multiwavelength studies in the direction of TeV shells can provide important 
empirical evidence to confirm their nature as SNRs and to establish the emission mechanism. 
Among the 14 TeV shells reported in the TeVCat\footnote{\url{http://tevcat2.uchicago.edu/}}, only 
HESS J1614$-$518 and HESS J1912$+$101 lack a confident counterpart in another spectral band, and 
their nature remains unclear.
In this paper we present a radio and infrared study of the intriguing source HESS J1912$+$101.
This TeV shell has been widely studied since its discovery but has still not been confirmed as a SNR, and 
it is not clear whether the $\gamma$-ray emission is dominated by leptonic or hadronic processes. 

HESS J1912$+$101 was discovered by \citet{ahar08} as a new $\gamma$-ray source in the H.E.S.S. Galactic Plane Survey (HGPS).
The emission was identified as extended and fitted with a Gaussian profile, which yielded an intrinsic extension of $\sim 0.3^{\circ}$ 
(full width at half maximum) and a centroid located at RA: 19$^{\rm{h}}$ 12$^{\rm{m}}$ 49$^{\rm{s}}$, Dec. = +10\d 09\m 6\m\m. 
The pulsar PSR J1913$+$1011 was proposed as the most plausible source powering the $\gamma$-ray emission. Indeed, it is 
an energetic pulsar, with a spin-down power $\dot{E}=2.9\times 10^{36}\,\mathrm{erg\,s^{-1}}$ and a characteristic age 
$\tau_c= 1.7\times 10^5\,\mathrm{yr}$, which appears projected over the TeV emission, though $\sim 0.15^{\circ}$ offset
from the TeV centroid derived by the authors. 
The efficiency of conversion of the pulsar power into TeV luminosity is $0.5\%$, similar to other
TeV pulsar wind nebula (PWN)--pulsar associations. From the dispersion measure (DM) of $179\,\mathrm{cm^{-3}\,pc}$, 
a pulsar distance of $\sim 4.6\,\mathrm{kpc}$ was derived using the electron density model YMW16 of \citet[refer to the ATNF Pulsar Catalog; \citealt{Manchester05}]{yao17}.
 
Even though the presence of the pulsar PSR J1913$+$1011 supports a PWN origin for HESS J1912$+$101, the non-detection of a nebular
counterpart in other spectral ranges and the lack of an energy-dependent morphology in the TeV band are arguments against this scenario.  
Subsequent observations obtained with H.E.S.S. by \citet{2015Puehlhofer} , 
which increase the total exposure by a factor of 6 with respect to the HGPS,
revealed a shell-like morphology, with inner and outer radii of $0.28^{\circ}$ and $0.40^{\circ}$, respectively. 
Interestingly, PSR J1913$+$1011 appears located at the geometrical center of the TeV shell.
Based on the shell appearance in the TeV band, HESS J1912$+$1011 was classified as a SNR candidate, and 
\citet{gottschall17} suggest a hadronic origin for the $\gamma$ rays due to the lack of nonthermal X-ray emission.

Indirect evidence of the presence of a SNR coincident with HESS J1912$+$101 was provided by \citet{su2017}, 
who analyzed the distribution of the molecular and neutral gas in the region.
They find a couple of expanding shells of CO and HI to the west of the TeV emission, 
which may have been accelerated after the interaction of the SNR with 
interstellar material located at a systemic velocity of $\sim 60\,\mathrm{km\,s^{-1}}$. 
They conclude that this material is located at a near kinematic distance of $\sim 4.1\,\mathrm{kpc}$ and
estimate the age of the putative SNR to be $\sim 0.7 - 2.0 \times 10^5\,\mathrm{yr}$. 
Interestingly, both the distance and the age of the remnant are compatible with those of 
the pulsar PSR J1913$+$1011, favoring an association between them. 
\citet{reich19} performed a radio study of HESS J1912$+$101 and found a partial shell of polarized emission
at $6\,\mathrm{cm}$, which they attribute to the presence of a SNR.
Indeed, this arc-like structure is located to the east of the TeV emission, in a region free of clumpy molecular gas, where 
the ambient magnetic field can be efficiently compressed by the shocks of a SNR (and probably also by the stellar winds
of its progenitor star), leading to an increase in the synchrotron radiation. 
Even though the authors were not able to detect the emission in the intensity maps, they 
estimate a flux density between 2.5 and $5.0\,\mathrm{Jy}$ from the flux density of the polarized arc.   
Such a low surface brightness is expected in an evolved SNR.

Contrary to a SNR origin, and motivated by the detection of HESS J1912$+$101 by the HAWK Observatory, \citet{linden17} suggest that it 
could be a member of the ``TeV halo'' population. TeV halos are very extended $\gamma$-ray sources associated with
middle-aged pulsars and PWNe ($\gtrsim 100\,\mathrm{kyr}$), where the pulsar wind is no longer efficiently 
confined by the interstellar medium (ISM) or the SNR ejecta, allowing a fraction of relativistic electron to diffuse away from it.
As a consequence, a compact nonthermal nebula of a few parsecs, produced by electrons
confined inside the pulsar wind, is expected to be observed in either radio or X-ray bands. More extended emission (of some tens of parsecs), produced by electrons
that have escaped the pulsar wind and radiate via the inverse Compton mechanism, is detected in the TeV band \citep{linden17,linden19}. 

\citet{zhang20} report on the detection of GeV emission from HESS J1912$+$101 using the Large Area Telescope (LAT) on board the
\textit{Fermi} Gamma-ray Space Telescope.
They conclude that the larger extension of the GeV source with respect to the TeV emission is compatible with a PWN
scenario, as higher-energy particles are expected to cool faster through inverse Compton scattering, producing a more compact
source in the TeV band. They claim that this discovery favors the TeV halo scenario but 
is contradicted by the shell-like appearance of HESS J1912$+$101. 
The authors propose that either the shell morphology observed by HESS is not real (the size and morphology measured
by HAWK are indeed different) and that both GeV and TeV sources come from a PWN powered by PSR J1913$+$1011, 
or only the GeV source is a PWN and the TeV shell is the associated 
SNR, which is producing TeV emission through the hadronic mechanism.     
Recently, \citet{zeng21} claimed that a leptonic origin for the GeV emission is only possible if the source is
a few millennia old, which does not match the age of PSR J1913$+$1011 ($\sim 170\,\mathrm{kyr}$). 
On the contrary, they find that a hadronic model gives an acceptable fit for the GeV-TeV emission. 

Another pulsar is seen superimposed on HESS J1912$+$101, at $\sim 10^{\prime}$ from the TeV centroid. 
PSR J1913$+$1000 is an old ($\tau_c = 7.9 \times 10^5$)
and low-energy ($\dot{E}=1.1\times 10^{33}\,\mathrm{erg\,s^{-1}}$) radio pulsar, located at a DM distance of $7.7\,\mathrm{kpc}$
\citep{Manchester05}. Based on its low spin-down flux density, $\dot{E}/D^2 \sim 1.9 \times 10^{-12}\,\mathrm{erg\,s^{-1}\,cm^{-2}}$,
this pulsar is not energetic enough to power the TeV emission of HESS J1912$+$101, which has an energy flux 
of $8.6 \times 10^{-12}\,\mathrm{erg\,s^{-1}\,cm^{-2}}$ between 1 and $10\,\mathrm{TeV}$ \citep{HESS18imagenes}. 

In summary, the astrophysical origin of HESS J1912$+$101 remains unclear, and possible scenarios are a SNR, a PWN, or a TeV halo.
On the one hand, HESS J1912$+$101 could be an old SNR producing TeV emission from the interaction
between protons accelerated at the shock front and dense material of the surrounding ISM.
On the other hand, TeV emission may originate in a PWN or a TeV halo powered by the 
energetic pulsar PSR J1913$+$1011, whose position, distance, and age are compatible with those derived for the putative SNR. 
In order to investigate the nature of HESS J1912$+$101, we performed high quality radio continuum observations
using the Karl. G. Jansky Very Large Array (VLA)\footnote{The VLA is a facility operated by the 
National Radio Astronomy Observatory (NRAO).}. We aimed to detect the radio counterpart of the TeV shell and determine 
if PSR J1913$+$1011 is powering a PWN. We also analyzed the properties of the ISM by studying the distribution of the neutral and
molecular gas and the infrared emission.

\section{Observations}

\subsection{New radio continuum observations}

HESS J1912$+$101 was observed with the VLA in the L band ($1.5\,\mathrm{GHz}$) using the D array configuration (Project ID: 18A$-$093). 
We mapped the entire extension of the TeV emission using the mosaicking technique, with 12 pointings forming an hexagonal path.  
We also observed the pulsar PSR J1913$+$1011 with a single pointing in the C band ($6.0\,\mathrm{GHz}$). 
Raw data were calibrated using the VLA pipeline.

To obtain the image in the L band, we used a chain of de-convolutions of three consecutive images: the model of the first image was 
taken as an input to deconvolve the second image, and so forth. 
The robust was allowed to range from 0.0 to 0.5 in order to recover a higher flux from the extended sources we detect in the image (most of them are cataloged HII regions).
Due to the large size of the map, we used the multifrequency and multi-scale deconvolution algorithms to account for the 
non-coplanarity of the mosaic. 
The final image has a resolution of $40^{\prime\prime}.1 \times 31^{\prime\prime}.9$, an angle of $-4$\d7, and an 
effective noise of $1\,\mathrm{mJy\,beam^{-1}}$.
In the C band, we used a robust factor of 0.5, and the final resolution of the image is $10^{\prime\prime}.9 \times 9^{\prime\prime}.1$, 
with an angle of $-2$\d.2, and an effective noise of $10\,\mathrm{\mu Jy\,beam^{-1}}$. 
In Table \ref{Radio_observaciones_1912} we summarize the observations and the parameters of the final maps.

\begin{table}[h]
\centering
  \caption{Summary of the observations and parameters of the final images. }
    \begin{tabular}{ccc}
    \hline \hline
          & HESS J1912$+$101 & PSR J1913$+$1011 \\
    \hline
    Observation date & 2018 Sep 10 & 2018 Sep 2\\ 
    Array configuration & D & D \\
    Frequency band & L ($1.5\,\mathrm{GHz}$) & C ($6\,\mathrm{GHz}$) \\
    Bandwidth & $1\,\mathrm{GHz}$ & $4\,\mathrm{GHz}$ \\
    Spectral windows & 16 ($64\,\mathrm{MHz}$) & 32 ($128\,\mathrm{MHz}$)\\
    Integration time & 9 min & 30 min \\
    Primary calibrator & 3C~48 & 3C~48 \\
    Secondary calibrator & J1859$+$1259 & J1922$+$1530 \\
    Map size (diameter) &  $1.4^{\circ}$    & $0.17^{\circ}$   \\
    Angular resolution & $40^{\prime\prime}.1\times 31^{\prime\prime}.9$ &  $10^{\prime\prime}.9 \times 9^{\prime\prime}.1$ \\
    Sensitivity (rms) &  $1\,\mathrm{mJy\,beam^{-1}}$ & $10\,\mathrm{\mu Jy\,beam^{-1}}$ \\
    \hline
    \end{tabular}
\tablefoot{In the L band, ``integration time'' refers to the total on-source time for each pointing of the mosaic. }    
\label{Radio_observaciones_1912}
\end{table}

\subsection{Additional data}

The ISM around HESS J1912$+$101 was investigated using the neutral hydrogen (HI) data extracted from the Very Large Array Galactic Plane Survey (VGPS; \citealt{stil06}), which maps the HI 21 cm line with angular and velocity resolutions of 1\m~and $0.8\,\mathrm{km\,s^{-1}}$, respectively. The mean noise ({rms}) per channel is $\sigma_{rms}=2\,\mathrm{K}$.
We also used the Galactic Ring Survey (GRS; \citealt{jack06}) to explore the distribution of the molecular gas. 
The survey maps the first Galactic quadrant in the $^{13}$CO J = 1--0 line with angular and spectral velocity resolutions of 46\m\m~and 
$0.2\,\mathrm{km\,s^{-1}}$, respectively. The mean noise (rms) per channel is $\sigma_{rms}=0.13\,\mathrm{K}$.
Additionally, we used the infrared images at $8\,\mathrm{\mu m}$ and $24\,\mathrm{\mu m}$ from the 
Galactic Legacy Infrared Midplane Survey Extraordinaire (GLIMPSE) performed with the
\textit{Spitzer} Space Telescope, which have angular resolutions of $1^{\prime\prime}.9$ and $6^{\prime\prime}.0$, respectively.   

\section{Results}

\subsection{The new VLA images} 
\label{radio_img}

\begin{figure*}[h]
\centering
\includegraphics[width=0.75\textwidth]{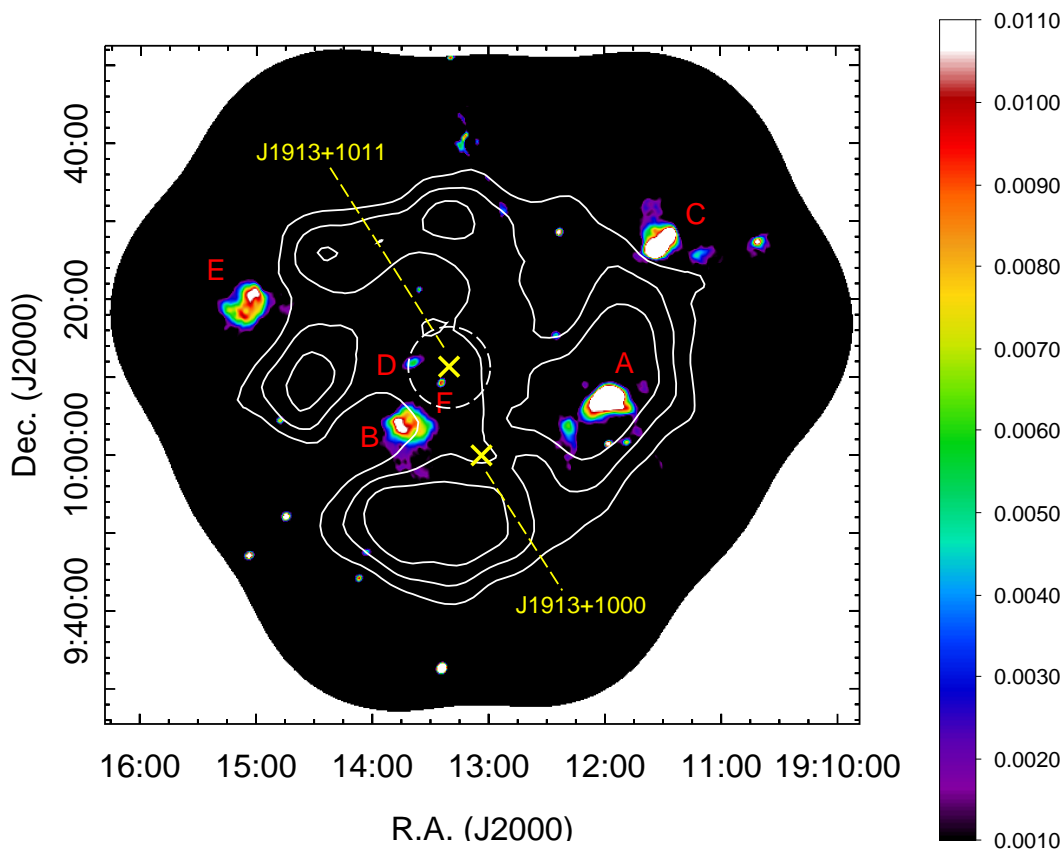}
\caption{Radio continuum image at $1.5\,\mathrm{GHz}$, overlaid with TeV emission (white contours). 
The yellow crosses indicate the positions of PSR J1913$+$1011 and PSR J1913$+$1000, and the white circle is the
region observed at $6.0\,\mathrm{GHz}$. The radio sources analyzed in the text are labeled with red capital letters. 
The angular resolution is $40^{\prime\prime}.1 \times 31^{\prime\prime}.9 $, PA=$-4$\d.7, and the 
noise is $1\,\mathrm{mJy\,beam^{-1}}$. The intensity color scale is expressed in $\mathrm{Jy\,beam^{-1}}$. }
\label{Radio_hess_1912}
\end{figure*}

Figure \ref{Radio_hess_1912} presents the new radio continuum image at $1.5\,\mathrm{GHz}$, which shows the region observed at $6.0\,\mathrm{GHz}$ toward PSR J1913$+$1011.
TeV emission from HESS J1912$+$101 is also shown\footnote{Retrieved from \url{https://www.mpi-hd.mpg.de/hfm/HESS/hgps/}}.
We do not detect any radio emission that could be attributed to the shell of a SNR.
The emissions with a stripe-like appearance to the north are probably interference patterns caused by radio sources
located outside the field of view of the mosaic but still detected by the side lobes of the antennas.

Several bright sources are present in the field; most of them are confirmed or candidate HII regions identified
in former radio surveys (references are shown in Table \ref{tabla_HIIradio}).
Radio source {\it A} overlaps the western TeV emission and corresponds to the HII region U44.26$+$0.10.
Radio source {\it B} is located at the border of HESS J1912$+$101 and 
is associated with the HII region G044.427$-$0.317.
Source {\it C} is formed by two distinct radio sources, which do not overlap the TeV emission and 
are associated with the HII regions G044.501$+$0.335 and G044.518$+$0.379.
Source {\it D} is the radio HII region G044.528$-$0.234.
Radio emission from source {\it E} corresponds to the HII region D44.79$-$0.49. 
Finally, we have marked the position of radio source {\it F}, identified as NVSS J191324$+$100916
in the NRAO VLA Sky Survey (NVSS) at $1.4\,\mathrm{GHz}$. No radio or infrared HII region 
is cataloged in the direction of this radio source, but we do detect such a region in our new $6.0\,\mathrm{GHz}$ image (see below). 
In Table \ref{tabla_HIIradio} we report the list of radio HII regions, the associated
infrared sources from the Wide-field Infrared Survey Explorer (WISE) 
Catalog of Galactic HII Regions, and their systemic velocities obtained
from radio recombination and/or molecular line observations.
For all the HII regions with positive velocities, the distance ambiguity was solved in favor of the far distance
(see the references). The kinematic distances were
obtained using the Galactic rotation model of \citet[hereafter Reid14]{reid2014}\footnote{To convert from velocity to distance,
we used the online tool developed by \citet{wenger18} and available at \url{https://www.treywenger.com/kd/index.php}.}.  

\begin{table*}[h]
\centering
\caption{Radio HII regions in the direction of HESS J1912$+$101.}
\begin{tabular}{clcccc}
\hline \hline
Label & Name & WISE  &$v$ ($\mathrm{km\,s^{-1}}$) & $D$ (kpc) & Ref. \\
\hline
A & U44.26$+$0.10     & G044.257$+$00.095 & 59.6    & 7.8  & 1,2   \\ 
B & G044.427$-$0.317  & G044.427$-$00.317 & 61.1    & 7.7  & 3,4     \\
C & G044.501+0.335    & G044.501$+$00.332 & $-43.0$ & 15.5 & 3,4     \\
  & G044.518+0.379    & G044.521$+$00.385 & $-49.7$ & 16.2 & 3,4     \\
D & G044.528$-$0.234  & G044.552$-$00.239 & 58.4    & 7.9  & 3,4 \\
E & D44.79$-$0.49     & G044.811$-$00.492 & 44.8    & 8.9  & 1,2 \\
   \hline
   \end{tabular}
\tablefoot{The ``Label' column refers to Fig. \ref{Radio_hess_1912}. For sources A and E, the ``Name'' of the radio HII regions is taken from reference (2), while for sources B, C, and D it is taken from reference (3). The ``WISE'' column shows
the associated infrared source from the WISE Catalog of Galactic HII Regions \citep{Anderson14}.  
The column ``$v$'' reports the systemic velocity reported in the references, and ``$D$'' is the corresponding kinematic distance 
using the Reid14 Galactic rotation model.
Ref.: (1) \citet{lockman89}, (2) \citet{anderson09a}, (3) \citet{anderson11}, and (4) \citet{anderson12}.}
\label{tabla_HIIradio}
\end{table*}

\begin{figure*}[ht!]
\centering
\includegraphics[width=0.81\textwidth]{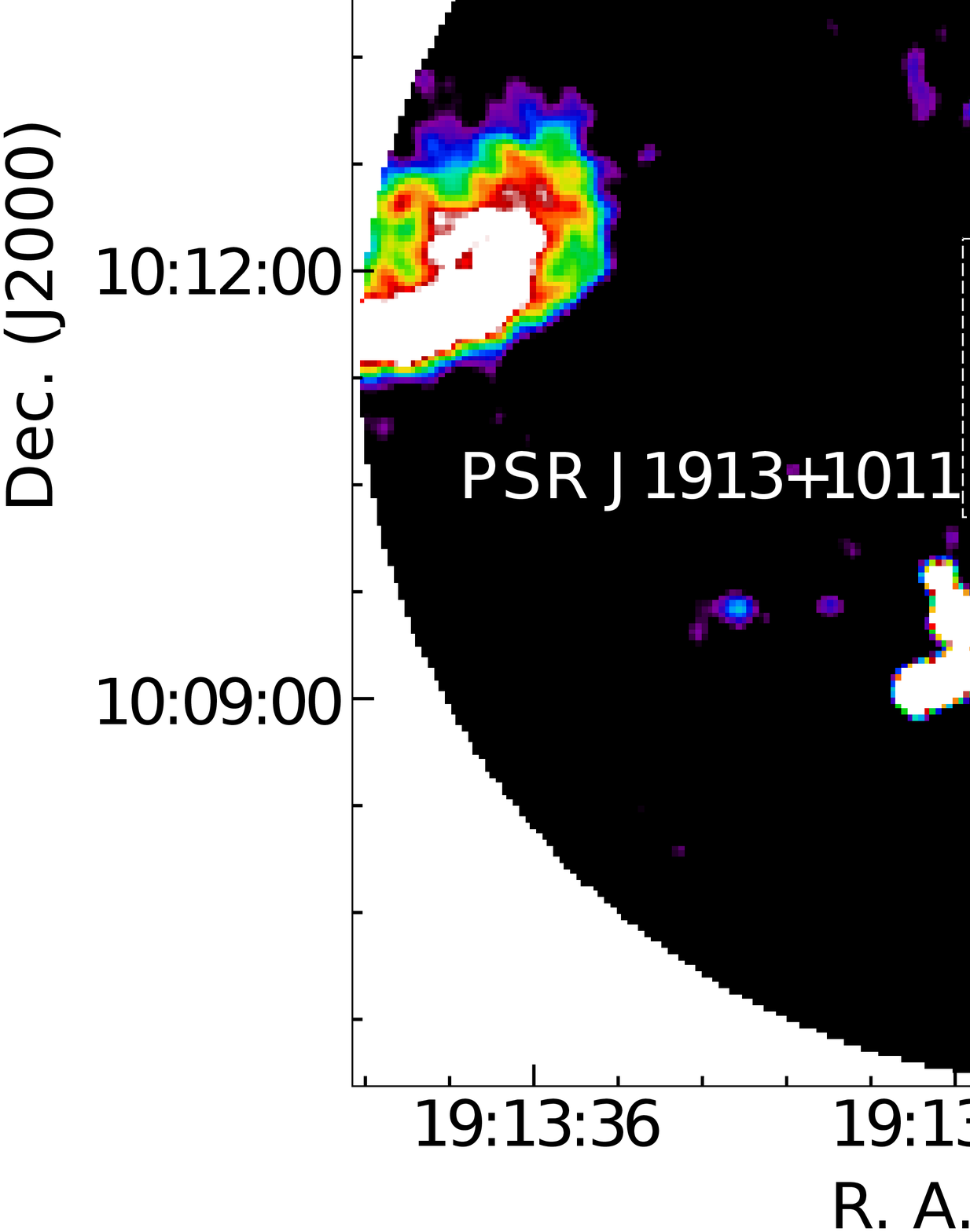}
\caption{Radio emission at $6.0\,\mathrm{GHz}$ in the direction of the pulsar PSR J1913$+$1011, 
whose position is indicated with a white cross. 
The final resolution is 10\m\m.9 $\times $9\m\m.1, and the effective noise level is 10 $\mu$Jy beam$^{-1}$. 
The upper-right panel shows a zoomed-in view of the radio emission toward the pulsar, and the bottom-right panel
shows a two-color image of the infrared emission ($8\,\mathrm{\mu m}$ in red and $24\,\mathrm{\mu m}$ in green).
The intensity color scale is expressed in $\mathrm{Jy\,beam^{-1}}$.} 
\label{Radio_pulsar_1913}
\end{figure*}

Figure \ref{Radio_pulsar_1913} shows the new image at $6.0\,\mathrm{GHz}$ around PSR J1913$+$1011. 
Two bright sources stand out in the image. The first, the bright emission to the east at the border of the observed field, 
corresponds to the HII region G044.528$-$0.234 (radio source {\it D} of Fig. \ref{Radio_hess_1912}). 
The second is a bright source to the southwest of the image center, at the position of the radio source {\it F} detected
in the $1.5\,\mathrm{GHz}$ image. In our image at $6.0\,\mathrm{GHz}$, this bright source 
shows a complex morphology formed by two distinct radio spots.

Interestingly, there is faint diffuse emission at $6.0\,\mathrm{GHz}$ in the direction of the pulsar. 
To highlight this emission, Fig. \ref{Radio_pulsar_1913} (top right) displays a zoomed-in view
of the region. The emission is detected at above three times the map noise; 
it has an extension given by the semi-axis of $\sim 15^{\prime\prime} \times 22^{\prime\prime}$ and a flux of $0.17\,\mathrm{mJy}$.
Based on the positional coincidence with PSR J1913$+$1011, this emission may represent a PWN powered by the pulsar. 
To further investigate this possibility, 
in the bottom-right panel of Fig. \ref{Radio_pulsar_1913} we display the infrared emission of the enlarged region. 
There is a point source at $8.0\,\mathrm{\mu m}$ coincident with the radio emission, but no diffuse emission is detected
in either the $8.0$ or the $24.0\,\mathrm{\mu m}$ bands, as would be expected for a HII region.
The lack of an extended infrared counterpart could indicate that the radio emission 
is nonthermal, representing a synchrotron nebula powered by PSR J1913$+$1011.
In order to confirm the PWN nature of the $6.0\,\mathrm{GHz}$ emission, a spectral and polarimetry radio study needs to be carried out.

\subsection{Molecular and neutral gas distribution}
\label{ISM}

\begin{figure*}[h]
\centering
\includegraphics[width=0.8\textwidth]{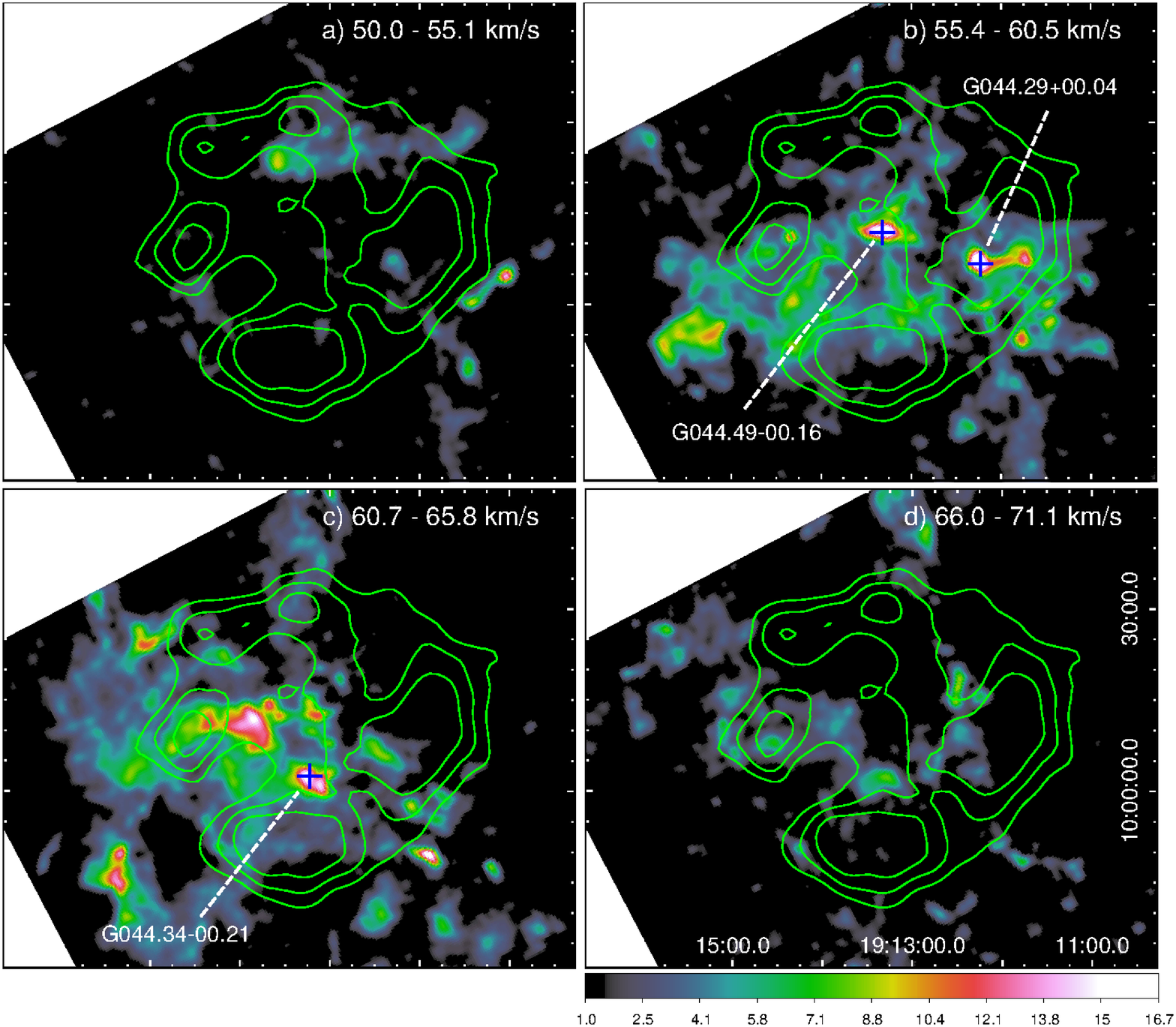}
\caption{Maps of the $^{13}$CO J=1--0 emission integrated every $5\,\mathrm{K\,km\,s^{-1}}$. The contours are the TeV emission, and the blue
crosses indicate the centroid of the three 
molecular clouds identified in the GRS at $\sim 60\,\mathrm{km\,s^{-1}}$.
The rms noise of each panel is $\sigma_{rms}\times \Delta v \times \sqrt{N_{chan}} \sim 0.3\,\mathrm{K\,km\,s^{-1}}$, 
where $\sigma_{rms}$ and $\Delta v$ are the mean rms noise per channel and the velocity resolution of the GRS, respectively,  
and $N_{chan}=24$ is the number of integrated channels. The color scale is expressed in $\mathrm{K\,km\,s^{-1}}$. Equatorial (J2000)
coordinates are indicated in panel {\it d}.}
\label{fig_CO2}
\end{figure*}

We analyzed the large-scale distribution of the molecular gas in the direction of HESS J1912$+$101
by inspecting the $^{13}$CO J=1--0 cube in the whole velocity range.
We find molecular gas overlapping the TeV emission at $\sim 60\,\mathrm{km\,s^{-1}}$.
In Fig. \ref{fig_CO2} we show the CO emission integrated every $5\,\mathrm{km\,s^{-1}}$ between 50 and 
$70\,\mathrm{km\,s^{-1}}$. This large molecular complex extends in the west-to-east direction, overlapping
the western and eastern regions of the TeV shell and only marginally overlapping the 
southern peak of TeV emission. 
We note that this molecular complex corresponds to the ``quiescent molecular gas'' identified by \citet[refer to their Fig. 4]{su2017}. They place it at the near kinematic distance because they find HI dips in the direction
of the CO peaks, which indicates that the clouds are absorbing against background gas at the far kinematic distance. 
We note, however, that \citet{rathborne09} identified three molecular clouds in the GRS around 
$\sim 60\,\mathrm{km\,s^{-1}}$: GRSMC G044.49$-$00.16 ($v=60.3\,\mathrm{km\,s^{-1}}$), GRSMC G044.29$-$00.04 ($v=56.9\,\mathrm{km\,s^{-1}}$),
and GRSMC G044.34$-$00.21 ($v=65.0\,\mathrm{km\,s^{-1}}$).
We have indicated the centroid of these clouds in panels {\it b} and {\it c} of Fig. \ref{fig_CO2}.
According to \citet{roman-duval09}, these clouds are located at their
far kinematic distances, which correspond to $\sim 7.7$, $\sim 8.1$, and $\sim 7.2\,\mathrm{kpc}$, respectively, according to the 
Reid14 Galactic rotation model.

\begin{figure*}[h]
\centering
\includegraphics[width=\textwidth]{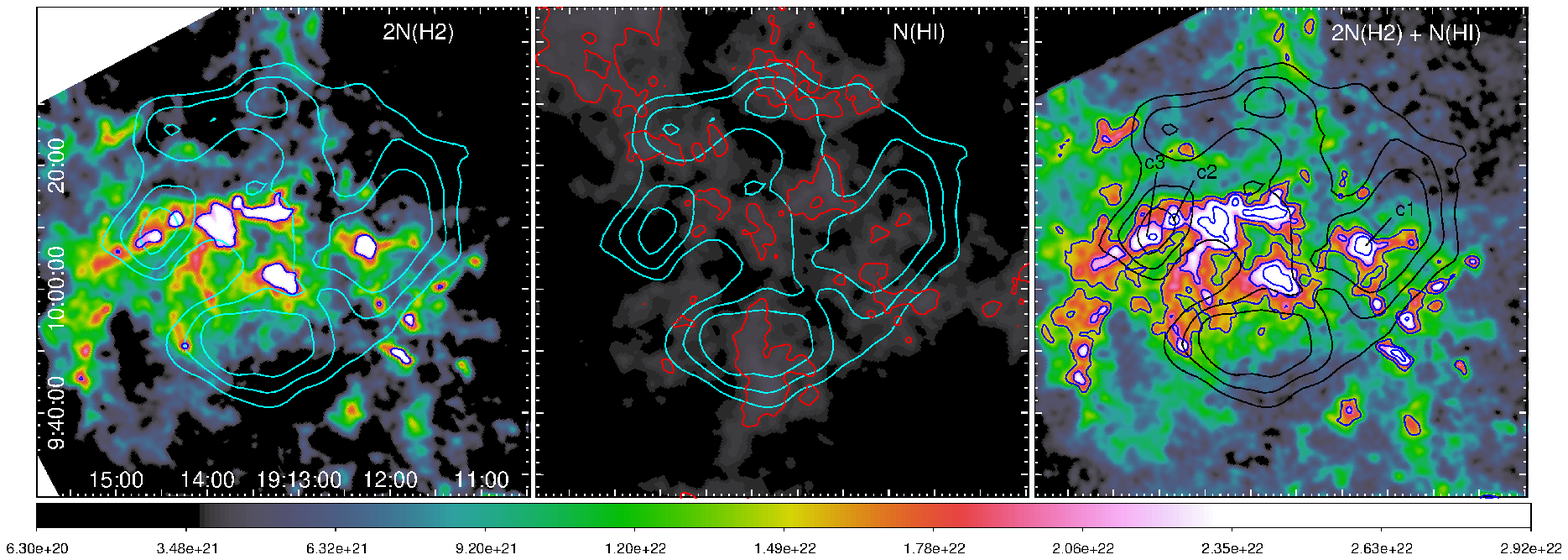}
\caption{Proton column density in the velocity interval between 50 and $70\,\mathrm{km\,s^{-1}}$. 
We show the contributions from the molecular gas (left panel, contour level $2\times 10^{22}\mathrm{cm^{-2}}$, 
$\sigma_{rms} = 2.2\times 10^{20}\,\mathrm{cm^{-2}}$) and the neutral gas (middle panel, contour level $4\times 10^{21}\mathrm{cm^{-2}}$, 
$\sigma_{rms} = 1.2\times 10^{19}\,\mathrm{cm^{-2}}$). The total column density is shown in the right panel
(contour levels 1.5, 2.0, and $2.7\times 10^{22}\mathrm{cm^{-2}}$). The color scale is expressed in $\mathrm{cm^{-2}}$. Equatorial (J2000)
coordinates are shown in the left panel.}
\label{1912_NH}
\end{figure*}

Regarding the neutral gas, we did not find any large-scale structure that may have been produced by a SNR, such as 
a cavity or a shell. In order to analyze the possible hadronic origin of HESS J1912$+$101, we obtained the proton distribution
of the ISM in the region. We considered the contributions from both molecular and neutral gas in the velocity
interval between 50 and $70\,\mathrm{km\,s^{-1}}$. The total proton column density is $N(H)= 2N(H_2)+ N(HI)$, where
$N(H_2)$ and $N(HI)$ are the molecular and neutral hydrogen column densities, respectively.  
For $N(H_2)$, we used Eqs. 1 and 2 of \citet{petriella13}, assuming $T_{ex}=10\,\mathrm{K}$, $T_b = 2.7\,\mathrm{K}$, 
and a relative abundance $N(H_2)=5\times 10^5 N(^{13}CO)$. Thus, $N(H_2)=5.1 \times 10^{20}\int T_B^{CO}\,dv\,\mathrm{cm^{-2}}$, 
where $T_B^{CO}$ is the brightness temperature of the $^{13}$CO J=1--0. 
For $N(HI)$, we used Eq. 4 of \citet{petriella13}, namely, $N(HI)=1.8 \times 10^{18}\int T_B^{HI}\,dv\,\mathrm{cm^{-2}}$,
where $T_B^{HI}$ is the brightness temperature of the HI emission. 
In Fig. \ref{1912_NH} we show the total proton column density (right panel)
obtained from the contributions of both molecular and neutral gas between 50 and $70\,\mathrm{km\,s^{-1}}$.
The CO image used to obtain $N(H_2)$ (left panel) was transformed to match the geometry of the HI image (middle panel).   

In Fig. \ref{1912_NH} we see that most of the ambient matter is associated with the molecular component.  
There is no clear correlation between the distribution of the ISM and the shell appearance
of HESS H1912$+$101. The most striking coincidence is observed in the western and eastern 
sides of the TeV shell. We have identified three molecular clumps ($c1$ to the west and $c2$ and $c3$ to the east), whose positions
match the $\gamma$-ray emission. We note that $c1$ corresponds to the molecular cloud GRSMC G044.29$-$00.04.
Based on the disagreement between \citet{su2017} and \citet{roman-duval09} regarding the distance to the molecular 
gas centered at $60\,\mathrm{km\,s^{-1}}$, we estimated the physical parameters of the ambient
gas considering near and far distances of $\sim 4.1$ and $\sim 7.8\,\mathrm{kpc}$, respectively, obtained from the 
Reid14 Galactic rotation model.
We derived the total column density of the three clumps by integrating the emission within the $2.0\times 10^{22}\mathrm{cm^{-2}}$
contour level of the right panel of Fig. \ref{1912_NH}. We assumed the emission comes from spherical clumps with
radii of 2.7, 2.0, and 2.5 arcmin for $c1$, $c2$, and $c3$, respectively. Masses and densities are reported in 
Table \ref{tabla_nubes}. For the masses, we took a relative He abundance of 25$\%$.     

\begin{table}[h]
\centering
  \caption{Physical parameters of the clumps.}
    \begin{tabular}{cccccccc}
    \hline \hline
Clump & $N(H)$  &\multicolumn{2}{c}{ $n$ ($\times 10^3\,\mathrm{cm^{-3}}$)} & \multicolumn{2}{c}{$M$ ($\times 10^3\,M_\odot$)} \\
      &  ($\times 10^{22}\,\mathrm{cm^{-2}}$)            & 4.1 kpc & 7.8 kpc                                     & 4.1 kpc & 7.8 kpc\\
    \hline
{\it c1} &  2.4 & 1.8 & 1.0 & 8.7 & 31.6\\ 
{\it c2} &  2.3 & 2.4 & 1.3 & 4.4 & 15.9 \\
{\it c3} &  2.2 & 1.8 & 0.9 & 6.7 & 24.4 \\
\hline
   \end{tabular}
   \label{tabla_nubes}
\end{table}

\subsection{Infrared emission}
\label{IR}

We constructed a two-color image of the infrared emission in the direction of HESS J1912$+$101 using the $8.0\,\mathrm{\mu m}$
and $24.0\,\mathrm{\mu m}$ images from the Infrared Array Camera (IRAC) and the Multiband Imaging
Photometer (MIPS) on board {\it Spitzer}, respectively.
There is bright emission in the western direction of the TeV shell. A zoomed-in image of this area is shown in Fig. \ref{fig_IR2}.
We have included the positions of the infrared HII regions (both confirmed and candidate) 
from the WISE Catalog of Galactic HII Regions, 
the \textit{Spitzer} dark clouds \citep[SDCs;][]{peretto09a}, and the $6.7\,\mathrm{GHz}$ methanol masers \citep{pandian07}.
The parameters of these sources are shown in Table \ref{tablaIR}. We also show contours of the 
$^{13}$CO emission integrated in the $53-60\,\mathrm{km\,s^{-1}}$ velocity range.

\begin{figure}[h]
\centering
\includegraphics[width=0.5\textwidth]{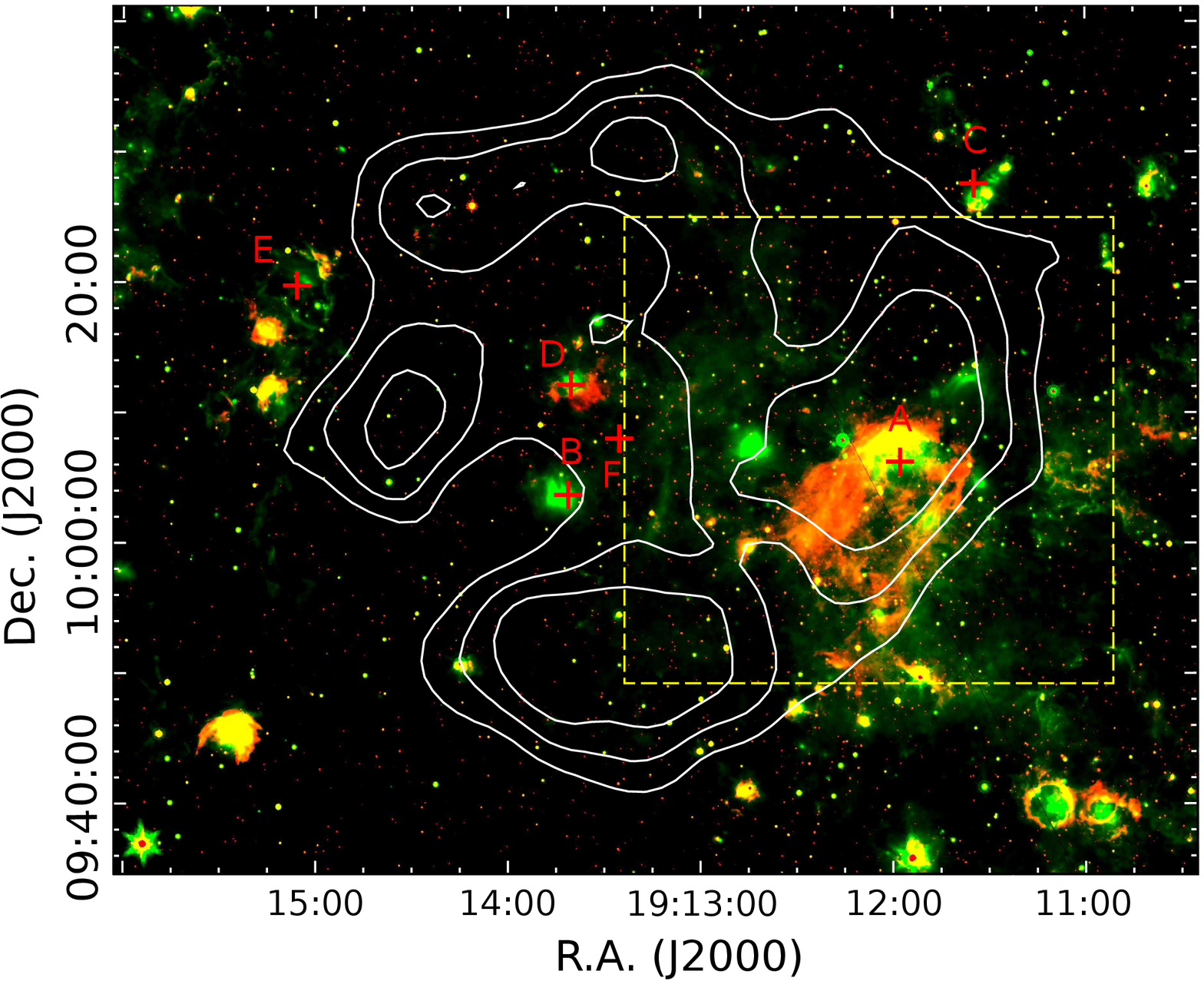}
\caption{Two-color image of the infrared emission toward HESS J1912$+$101: $8.0\,\mathrm{\mu m}$ (in red) and $24.0\,\mathrm{\mu m}$ (in green). The white contours are the TeV emission, and the yellow rectangle is a zoomed-in view of the area shown in Fig. \ref{fig_IR2}.
Red crosses are the radio sources of Fig. \ref{Radio_hess_1912}.}
\label{fig_IR1}
\end{figure}

\begin{figure}[h]
\centering
\includegraphics[width=0.5\textwidth]{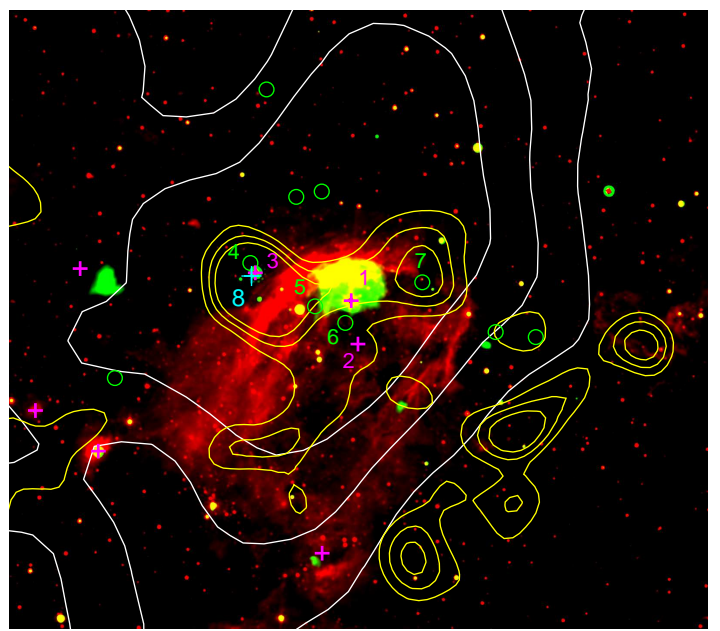}
\caption{Two-color image of the region indicated by the rectangle of Fig. \ref{fig_IR1}:
$8.0\,\mathrm{\mu m}$ (in red), $24.0\,\mathrm{\mu m}$ (in green), and TeV emission (white contours). 
The yellow contours represent the $^{13}$CO emission integrated in the $53-60\,\mathrm{km\,s^{-1}}$ velocity interval (contour levels:
6.5, 8.3, and $10.4\,\mathrm{K\,km\,s^{-1}}$). Magenta crosses indicate the position of the infrared HII regions, 
green circles the position of the infrared-dark clouds, and 
the cyan cross the methanol maser. The list of sources is given in Table \ref{tablaIR}.}
\label{fig_IR2}
\end{figure}

\begin{table*}[ht]
\centering
    \caption{Sources labeled in Fig. \ref{fig_IR2}. Refer to the text for the references.}
    \begin{tabular}{ccccc}
    \hline \hline
Label & Name  & Source type & Velocity ($\mathrm{km\,s^{-1}}$)  & Distance (kpc) \\
    \hline
1 & G044.257$+$00.095& HII region & 59.6 & 7.8 \\ 
2 & G044.224$+$00.085& HII region & -    & - \\ 
3 & G044.310$+$00.040& HII region & 57.6 & 3.9 / 8.0 \\ 
4 & SDC 3717 & Dark cloud & - & - \\ 
5 & SDC 3714 & Dark cloud & - & - \\ 
6 & SDC 3712 & Dark cloud & - & - \\ 
7 & SDC 3713 & Dark cloud & - & - \\ 
8 & G44.31+0.04 & Methanol maser & 55.7  & 3.8 / 8.2 \\ 
\hline
\end{tabular}
\tablefoot{The distances are derived from the systemic velocity using the Reid14 Galactic rotation model. 
For sources \#3 and \#8, near and far distances are reported because the ambiguity is not solved in the literature.}
\label{tablaIR} 
\end{table*}

The $8.0\,\mathrm{\mu m}$ band emission clearly reveals the presence of the infrared dust bubble (IRDB) N91. 
These types of bubbles are usually detected around HII regions as shells in specific bands of the mid-infrared, due to the excitation of polycyclic aromatic hydrocarbons in the photodissociation region \citep{church06}. 
Radio continuum and mid- and far-infrared continuum emission is commonly 
found in the interior of IRDBs. The former is produced by the thermal radio emission of the HII region, while the latter is 
produced by hot gas in the ionized gas region, near the excitation star(s), and is commonly detected in the $24.0\,\mathrm{\mu m}$ band
of \textit{Spitzer}-MIPS. 
Two WISE HII regions are cataloged inside N91. One of them, G044.224+0.085 (source \#2), is a radio-quiet HII region. The other one is 
G044.257$+$0.095 (source \#1), which has been identified as the infrared counterpart of the radio HII region U44.26$+$0.10 
(refer to Sect. \ref{radio_img}). From the systemic velocity of $59.6\,\mathrm{km\,s^{-1}}$ and using 
the HI emission/absorption method to solve the distance ambiguity, 
\citet{anderson09b} conclude that U44.26$+$0.10 is located at the far kinematic distance. 
According to the Reid14 Galactic rotation model, this corresponds to $7.8\,\mathrm{kpc}$.

There is a third infrared HII region in the this direction: G044.310$+$00.040 (source \#3). 
It is a candidate WISE HII region, whose position coincides with an infrared-dark cloud (source \#4) and the $6.7\,\mathrm{GHz}$ methanol
maser G44.31$+$0.04 \citep[source \#8;][]{pandian07}.
The Bolocam $1.1\,\mathrm{mm}$ continuum source BGPS G044.307$+$00.043 is also 
associated with it \citep{schlin11}. \citet{schlin11} 
report the detection of HCO$^{+}$ J=3--2 at a velocity of $57.6\,\mathrm{km\,s^{-1}}$, although they
cannot solve the ambiguity between near and far kinematic distances. 
In the direction of the methanol maser, \citet{pandian12} detect ammonia (NH$_3$) inversion lines
at a velocity of $56.2\,\mathrm{km\,s^{-1}}$. 

To summarize, we find several infrared sources coincident with the western TeV shell. 
The IRDB N91 (likely associated with the radio HII region U44.26$+$0.10)
is probably interacting with the ISM, as revealed by the presence of molecular gas over its infrared shell. 
This molecular material harbors a radio HII region, an infrared-dark cloud, continuum dust emission at $1.1\,\mathrm{mm}$,
a methanol maser, and HCO$^{+}$ and NH$_3$ line emission. In the following section we show that these features are evidence
of a massive star-forming region, and we analyze the probable association with HESS J1912$+$101.

\section{Discussion}
\label{disc}

At present, two sources have been identified as possible particle accelerators in the direction of HESS J1912$+$101. 
The first, the pulsar PSR J1913$+$1011, may be powering an energetic wind and producing GeV and/or TeV emission 
through a leptonic mechanism. The second, a SNR, may be producing
$\gamma$ rays of hadronic origin as a consequence of the interaction with the surrounding ISM.
There is only indirect evidence of the presence of this SNR, but \citet{su2017} claim that it may be associated with PSR J1913$+$1011, 
based on the inferred age ($0.7-2.0\times 10^5\,\mathrm{yr}$) and distance ($4.1\,\mathrm{kpc}$), which are similar to those of the pulsar 
($1.7\times 10^5\,\mathrm{yr}$ and $4.6\,\mathrm{kpc}$).

We do not find evidence of the SNR in our new image at $1.5\,\mathrm{GHz}$. This is consistent
with the claim that the putative remnant is $\sim 10^5\,\mathrm{yr}$ old and therefore that the synchrotron radio emission has faded
below the sensitivity of the observations. 
We find faint diffuse emission around PSR J1913$+$1011 at $6.0\,\mathrm{GHz}$. Based on the position of this emission
and the lack of an infrared counterpart, we suggest that it might represent a radio PWN powered by the pulsar.
To explain the non-detection of the putative PWN in our image at $1.5\,\mathrm{GHz}$, we derived  
the expected flux ($F_\nu \propto \nu^{-\alpha} $) in this spectral band from the relation 
$F_{1.5}=F_{6.0}(1.5/6.0)^{-\alpha}$, where $F_{6.0}=0.17\,\mathrm{mJy}$ is the flux at $6.0\,\mathrm{GHz}$. 
Taking a typical radio spectral index for a PWN between 0.0 and 0.3 \citep{gaensler06} and expressing the area of the diffuse emission in terms of the beam at $1.5\,\mathrm{GHz}$ (see Table \ref{Radio_observaciones_1912}),
we obtain a flux density $S_{1.5}\sim 0.7 - 1.0\,\mathrm{mJy\,beam^{-1}}$. 
The expected flux density at $1.5\,\mathrm{GHz}$ is on the order of the sensitivity of the map at $1.5\,\mathrm{GHz}$ 
(Fig. \ref{Radio_hess_1912}), which explains the non-detection.

For the pulsar distance of $4.6\,\mathrm{kpc}$, the putative PWN detected at $6.0\,\mathrm{GHz}$ has a radius  
$R_{pwn} \sim 0.5\,\mathrm{pc}$, which is considerably smaller than the TeV radius obtained by HESS ($R_{TeV}\sim 20\,\mathrm{pc}$, from
the 1 $\sigma$ radius of $0.25^{\circ}$ reported by \citealt{HESS18imagenes}). 
Interestingly, we obtain $R_{TeV}/R_{pwn} \sim 40$, similar to other evolved 
PWN systems such as Geminga, the most likely TeV halo candidate known to date\footnote{It is worth noting that for Geminga
the PWN extent reported by \citet{giacinti20} was obtained from the X-ray emission, implying $R_{TeV}/R_{pwn} \sim 100$.} 
\citep{giacinti20}. In a TeV halo scenario, the radio emission at $6.0,\mathrm{GHz}$ would 
represent the synchrotron PWN produced by electrons confined 
in the wind of the middle-aged pulsar PSR J1913$+$1011, while HESS J1912$+$101 would be the associated TeV halo powered
by electrons that have escaped the pulsar wind. To further investigate this scenario, we estimated the electron density within the TeV region  
as $\epsilon_e = \dot{E}\tau_c V^{-1}$, where $\dot{E}$ and $\tau_c$ are the spin-down power and characteristic age of 
the pulsar, respectively, and $V$ the volume of the $\gamma$-ray source calculated for a radius of $20\,\mathrm{pc}$.
We obtain $\epsilon_e \sim 10\,\mathrm{eV\,cm^{-3}}$, considerably larger than the ISM electron energy
density ($\sim 0.1\,\mathrm{eV\,cm^{-3}}$), which suggests that the TeV emission is due to electrons that are still 
influenced by the pulsar (refer to the discussion of Fig. 2 of \citealt{giacinti20}). 
This indicates that HESS J1912$+$101 is the high-energy counterpart of the radio PWN rather than a TeV halo.



Despite the evidence supporting TeV emission being powered by PSR J1913$+$1011 (either as a TeV PWN 
or a TeV halo), and taking into account the fact that the majority of TeV shells have been associated
with SNRs detected in other spectral bands, the shell appearance of HESS J1912$+$101 points to a SNR origin.
The distance to the putative SNR was established by \citet{su2017} from the probable interaction with molecular material at a velocity of
$\sim 60\,\mathrm{km\,s^{-1}}$, for which they consider a near kinematic distance of $4.1\,\mathrm{kpc}$. 
However, several molecular clouds have been identified at $\sim 60\,\mathrm{km\,s^{-1}}$ in the GRS; \citet{roman-duval09} 
established that they are located at their far kinematic distances. 
Thus, it is not evident that all this molecular material is located at the near distance.
Indeed, if we consider a far kinematic distance of $\sim 7.8\,\mathrm{kpc}$, the SNR can be associated with
the pulsar PSR J1913$+$1000, which has a DM distance of $\sim 7.6\,\mathrm{kpc}$. Based on the characteristic age of
this pulsar ($\sim 7.9\times 10^5\,\mathrm{yr}$), we expect the radio emission from the shell to be undetected.    
Due to its low spin-down power, this pulsar would not contribute to the TeV emission from HESS J1912$+$101, and the $\gamma$-ray emission
would be entirely produced by the SNR. 

We investigated if the putative SNR could be producing TeV emission of hadronic origin considering the two 
possible distances. For this purpose, we obtained a crude estimate of the ambient density, $n_0$, needed 
to account for the TeV flux using the equation of \citet{Torres03}:
\begin{equation}
F_{\gamma} (> E) = 10^{-10} f_{\Gamma} E^{-\Gamma +1}_{\mathrm{TeV}}A,
\end{equation}
where the factor $f_{\Gamma}$ depends on the spectral index $\Gamma$, $E_{\mathrm{TeV}} = E/TeV$, and 
$A = \theta E_{51} D^{-2}_{kpc}n_0~\mathrm{ph\,cm^{-2}\,s^{-1}}$. 
Here, $\theta$ is the efficiency of conversion of the supernova energy into cosmic ray energy, 
$E_{51}$ is the mechanical energy of the supernova explosion in units of $10^{51}\,\mathrm{erg}$, $D_{kpc}$ is the distance in kpc, 
and the ambient density, $n_0$, is expressed in cm$^{-3}$.
We used $F_{\gamma}(\geq 1 \mathrm{TeV})~2.49\times 10^{-12}\,\mathrm{cm^{-2}\,s^{-1}}$ and the spectral index $\Gamma =2.56$ 
\citep{HESS18imagenes}, an acceleration efficiency for protons of $10\%$ \citep{dermer13}, 
and a canonical supernova explosion of $10^{51}\,\mathrm{erg}$. Doing so, we find the density required for the ISM to act as a target for
hadronic $\gamma$-ray emission to be $> 20\,\mathrm{cm^{-3}}$ (for a distance of 
$4.1\,\mathrm{kpc}$) and $> 80\,\mathrm{cm^{-3}}$ (for a distance of $7.8\,\mathrm{kpc}$). 
The densities of the molecular clumps $c1$, $c2$, and $c3$ are by far greater than $n_0$, 
for both near and far distances. Thus, all the clumps are dense enough to produce the observed TeV flux of HESS J1912$+$101.

Based on the good positional coincidence between the molecular clumps and the TeV emission in the
western and  eastern directions, we suggest that the hadronic mechanism could be responsible 
for (at least) part of the TeV radiation. However, two problems arise when considering the hadronic scenario: (i) the lack of ambient matter correlating with the northern and southern regions of the TeV shell and (ii) the lack of TeV emission at the center of HESS J1912$+$101, where we also detect dense molecular gas.
The molecular clumps at the center HESS J1912$+$101 have radii
of $\sim 3\,\mathrm{arcmin}$ ($\sim 4$ and $\sim 7\,\mathrm{pc}$, for a distance of 4.1 and $7.8\,\mathrm{kpc}$, respectively) 
and densities $\gtrsim 1000\,\mathrm{cm^{-3}}$, similar to $c1$, $c2$, and $c3$.
A rough estimate of the diffusion length, $l_{diff}$, of cosmic rays into this molecular material is obtained by combining
Eqs. 5 and 6 of \citet{sano19}. Considering the Bohm limit (turbulence factor $\eta=1$), we get 
$l_{diff} [\mathrm{pc}]\sim 0.3 (E / 10\,\mathrm{TeV})^{0.5} (n / 300\,\mathrm{cm^{-3}})^{-0.325} t_{\mathrm{snr}}$,
where $E$ is the energy of the cosmic rays, $n$ is the ambient density, and $t_{snr}$ is the age of the SNRs in kyr.
Considering that the putative SNR is several tens of kiloyears old (as supported by the non-detection of a radio shell and the probable association 
with either PSR J1913$+$1011 or PSR J1913$+$1000), for calculation purposes  we took $t_{snr}=100$.
In a hadronic scenario, $\gamma$-ray photons with energies between 1 and $10\,\mathrm{TeV}$ are produced by
cosmic ray protons accelerated to energies $E \sim 10 - 100\,\mathrm{TeV}$.
Using this energy range and $n=1000\,\mathrm{cm^{-3}}$, we obtain $l_{diff}\sim 2 - 7\,\mathrm{pc}$, similar
to the radii of the molecular clumps detected in the region.  
Thus, despite the high density of the molecular clumps, TeV protons 
could have penetrated into them throughout the long lifetime of the SNRs, and we should be able to detect
$\gamma$ rays of hadronic origin from these clumps, as we do for the clumps $c1$, $c2$, and $c3$. 
We conclude that it is unlikely that TeV $\gamma$ rays are produced entirely by hadronic processes after the 
interaction of a SNR's shock front and the ISM. 
Indeed, if all the molecular material at $\sim 60\,\mathrm{km\,s^{-1}}$ is located at the same distance, we do not find a reasonable
explanation as to why the molecular gas in the western and eastern regions of the shell acts as an efficient target for 
proton-proton collision but the dense gas at the center of HESS J1912$+$101 does not.

Our study has also revealed another possible particle accelerator in the direction of HESS J1912$+$101: a star-forming region
in the direction of the western TeV emission, revealed by the presence of several infrared and radio sources.
Young stars have been identified as potential candidates of $\gamma$ rays in the GeV and TeV regimes.
Particles can be accelerated to relativistic energies at the shocks of protostellar jets when they collide
with the surrounding ambient medium \citep{Bosch-Ramon10}. 
So far, evidence of individual protostars powering $\gamma$-ray emission is scarce. 
For instance, the Herbig-Haro objects HH 219 \citep{hewit12} and HH 80-81 \citep{yan22} have been identified as potential GeV sources.
Efficient particle acceleration in star-forming regions is also expected to arise from the combined action
of several protostars and young stellar clusters \citep{romero08}, or by the collective action of the stellar winds of young massive stars
and SNRs within super-bubbles \citep{aharonian19}. 

We find that the IRDB N91 appears projected onto the $\gamma$-ray emission. 
Both radio continuum and $24.0\,\mathrm{\mu m}$ emission are usually found within IRDBs. Therefore, the 
HII region U44.26$+$0.10 (radio source {\it A}) is likely associated with N91. 
The position of the radio emission, offset from the centroid of the infrared shell, 
can be explained by the expansion of the bubble in a nonhomogeneous medium. 
Indeed, the detection of molecular material to the north and east of the infrared shell (Fig. \ref{fig_IR2}) 
suggests that the IRDB may have encountered dense material in these directions but is expanding undisturbed in the other ones. It is worth noting that this material corresponds
to the molecular clump {\it c1} of Fig. \ref{1912_NH}, which is part of the molecular cloud GRSMC G044.29$+$00.04
identified by \citet{rathborne09}. From the cloud velocity of $56.9\,\mathrm{km\,s^{-1}}$, 
\citet{roman-duval09} solved the distance ambiguity using the HI self-absorption method and conclude that this molecular 
cloud is located at the far kinematic distance. On the other hand, the distance to U44.26$+$0.10 has been determined from 
the detection of a radio recombination line at a velocity of $59.6\,\mathrm{km\,s^{-1}}$. The 
distance ambiguity was solved in favor of the far distance, using the HI emission/absorption method \citep{anderson09b}.
Therefore, the HII region U44.26$+$0.10 and the molecular material are likely associated and are located at  
a distance of $\sim 7.8\,\mathrm{kpc}$. 

One of the effects of the expansion of HII regions in the ISM is the 
formation of new stars, and star-forming activity has been found around IRDBs (see \citealt{thompson12,petriella10,watson08}).
We have shown that N91 is probably associated with a couple of molecular clumps near its northern and eastern borders.    
This molecular material harbors the HII region candidate G44.310$+$0.040 (source \#3) 
and an infrared-dark cloud (source \#4). Infrared-dark clouds are considered an early stage in the formation
of massive stars \citep{rathborne07}. At this position, we find other traces of massive star formation: 
the presence of $6.7\,\mathrm{GHz}$ methanol maser emission,
which is mostly detected in massive star-forming regions \citep{paulson20}; 
the detection of HCO$^{+}$ line emission, which can reveal both 
molecular outflows \citep{tokuda14} and molecular envelope infall \citep{nagy2020} in massive protostars; and 
the detection of ammonia inversion lines, which trace high density gas in young protostellar cores \citep{vaisala14}.
Therefore, in the western direction of HESS J1912$+$101 and coincident with intense TeV emission, there is evidence of young stars with powerful stellar winds (as revealed by the HII region U44.26$+$0.10 and the IRDB N91)
and at least one massive protostar powering molecular outflows. This star-forming region could be contributing
to the TeV emission from HESS J1912$+$101. 

To summarize our major findings, we stress that HESS J1912$+$101 is located in a complex region, where we have identified three potential sources of $\gamma$ rays:
the pulsar PSR J1913$+$1011 (located at $\sim 4.6\,\mathrm{kpc}$), a SNR associated with molecular gas at 
$\sim 60\,\mathrm{km\,s^{-1}}$ (corresponding to kinematic distances of either $\sim 4.1\,\mathrm{kpc}$ or $\sim 7.8\,\mathrm{kpc}$), 
and a star-forming region at $\sim 7.8\,\mathrm{kpc}$. 
If the SNR is interacting with molecular material located at $\sim 4.1\,\mathrm{kpc}$, it can be associated with 
PSR J1913$+$1011, as suggested by previous work. In this case, the star-forming region
would be an unrelated source in the line of sight that is interacting with molecular material at the same velocity
but at the far distance. 
If the SNR is at the far distance, it could be associated with the pulsar PSR J1913$+$1000 (located at $\sim 7.6\,\mathrm{kpc}$),
but the contribution of the latter to the TeV emission would be negligible due to its low energy. 
In this case, the remnant and the star-forming region would be located at the same distance, suggesting that they might be associated.  
Indeed, the formation of the HII region U44.26$+$0.1 could have been triggered 
by the combined action of the SNR and the stellar
winds of its progenitor star. This scenario would represent an example of sequential star formation linking 
three generations of stars: (i) the progenitor star of the supernova, 
(ii) the ionizing stars of the HII region U44.26$+$0.10, and (iii) the protostar(s) embedded in the molecular clump at the 
border of the infrared shell of N91.

\section{Conclusions}

Our new radio continuum image at $1.5\,\mathrm{GHz}$ does not show evidence of a shell in correspondence 
with the TeV emission from HESS J1912$+$101, which would provide unquestionable evidence of a SNR.  
We found marginal evidence of nonthermal radio emission at $6.0\,\mathrm{GHz}$ associated with PSR J1913$+$1011, 
which indicates that the pulsar is probably powering a PWN. If the production of TeV emission is associated with the pulsar, 
further work is required to determine if it corresponds to either the very high-energy counterpart of the PWN or
its associated TeV halo. 

Based on the shell appearance and spectral properties in the TeV band, a SNR origin for HESS J1912$+$101 appears as the most
reliable hypothesis, but the underlying emission mechanism still remains elusive. 
Indeed, the large-scale distribution of the molecular and neutral gas shows good 
spatial correlation with the TeV shell only in the western and eastern directions.
Spatial correlation between TeV emission and molecular gas
is usually invoked as evidence of the hadronic mechanism. However, there is no dense material
in correspondence with the northern and southern parts of the shell. Furthermore, there is molecular gas near 
the center of the TeV shell, but this coincides with weak $\gamma$-ray emission.   
Thus, the distribution of the ambient gas does not favor a hadronic origin for the entire emission of HESS J1912$+$101.  

We conclude that part of the TeV emission from the shell may be hadronic, but we cannot rule out a leptonic process acting as well. Furthermore, two potential sources of high-energy particles that are present in the region
may also be contributing to the very high-energy emission: 
the pulsar PSR J913$+$1011 (which is probably powering a radio PWN) at the center of the shell and an active star-forming region to the west.
We plan to carry out a detailed study of this star-forming region in a forthcoming paper.  
In particular, a careful study of the molecular gas in the direction of the putative massive protostar(s)
will allow us to characterize the outflow activity and establish its contribution to the TeV emission from HESS J1912$+$101.

\section*{Acknowledgments}

L.D. is postdoctoral fellow of CONICET, Argentina. 
A.P. is member of the {\sl Carrera del Investigador Cient\'\i fico} of CONICET, Argentina. 
This work was partially supported by Argentina grants awarded by UBA (UBACyT) and ANPCyT.
The National Radio Astronomy Observatory is a facility of the National Science Foundation 
operated under cooperative agreement by Associated Universities, Inc.
This publication makes use of molecular line data from the Boston University-FCRAO Galactic Ring Survey (GRS). 
The GRS is a joint project of Boston University and Five College Radio Astronomy Observatory, funded by the National
Science Foundation under grants AST-9800334, AST-0098562, AST-0100793, AST-0228993, \& AST-0507657.

\bibliographystyle{aa}  
\bibliography{ref}
\IfFileExists{\jobname.bbl}{}
{\typeout{}
\typeout{****************************************************}
\typeout{****************************************************}
\typeout{** Please run "bibtex \jobname" to optain}
\typeout{** the bibliography and then re-run LaTeX}
\typeout{** twice to fix the references!}
\typeout{****************************************************}
\typeout{****************************************************}
\typeout{}
}

\label{lastpage}
\end{document}